%% file: proceedings_NuPhys18_arXiv_V2.tex
\documentclass[12pt]{article}
\pdfoutput=1
\usepackage[utf8]{inputenc}
\usepackage{amsmath}
\usepackage{amssymb}
\usepackage{units}
\usepackage{graphicx}
\usepackage{slashed}
\usepackage{array}
\usepackage[leftcaption,raggedright]{sidecap}
\usepackage{caption}
\usepackage[hidelinks]{hyperref}

\newcommand\pubnumber{NuPhys2018-Brivio}
\newcommand\pubdate{\today}

\def\NBI{Niels Bohr International Academy and Discovery Center\\
Niels Bohr Institute, University of Copenhagen, DK-2100 Copenhagen, Denmark}

\def\ITP{Institut f\"ur Theoretische Physik\\
Universit\"at Heidelberg, DE-69120 Heidelberg, Germany}

\def\support{\footnote{Work partially supported by the Villum Fonden, the Danish National Research Foundation under contract DNRF91,
the state of Baden-W\"urttemberg through bwHPC and the German Research Foundation (DFG) through grant no INST 39/963-1 FUGG(bwForCluster NEMO).
}}

\def\Title#1{\begin{center} {\Large #1 } \end{center}}
\def\Author#1{\begin{center}{ \sc #1} \end{center}}
\def\Address#1{\begin{center}{ \it #1} \end{center}}

\newcommand\pubblock{\rightline{\begin{tabular}{l} \pubnumber\\
         \pubdate  \end{tabular}}}
\newenvironment{Abstract}{\begin{quotation}  }{\end{quotation}}
\newenvironment{Presented}{\begin{quotation} \begin{center} 
             PRESENTED AT\end{center}\bigskip 
      \begin{center}\begin{large}}{\end{large}\end{center} \end{quotation}}

\input econfmacros.tex

\begin{document}
\begin{titlepage}
\pubblock

\vfill
\Title{The Neutrino Option}
\vfill
\Author{ Ilaria Brivio\support}
\Address{\NBI}
\Address{\ITP}
\vfill
\begin{Abstract}
The Neutrino Option is a scenario where the Higgs mass is generated at the same time as neutrino masses in the type-I seesaw model. This framework provides a dynamical origin for the scalar potential of the Standard Model and suggests a new approach to the hierarchy problem.
Here we review the preliminary analysis of Ref.~\cite{Brivio:2017dfq}, that showed the viability of this scenario, as well as the improved study of Ref.~\cite{Brivio:2018rzm}, that led to a better identification of the region of the parameter space where the Neutrino Option can be realized. 
We find that experimental constraints from both Higgs and neutrino physics can be accommodated introducing 2 heavy Majorana neutrinos with mass $M_1\simeq M_2\sim 0.5 - 10$ PeV and Yukawa couplings to the lepton doublet of order $ 10^{-4}-10^{-2}$, assuming that at the scale $M$ the classical Higgs potential is approximately conformal, with a quartic Higgs coupling $\lambda_0\sim 0.01-0.05$. Specifying the light neutrino mass ordering, the ratio $M_2/M_1$ or a given value of the top quark mass identifies narrower ranges for all the parameters.
Although no further signature of the Neutrino Option is generally predicted at the currently accessible energy scales, conformal UV completions have been proposed, that could be tested \eg\ via detection of gravitational waves. Leptogenesis can also be successfully realized in this scenario, that intriguingly ties together the breaking of the conformal and electroweak symmetries with the violation of lepton number. 

\end{Abstract}
\vfill
\begin{Presented}
NuPhys2018, Prospects in Neutrino Physics\\
Cavendish Conference Centre, London, UK,\\ December 19--21, 2018
\end{Presented}
\vfill
\end{titlepage}
\def\thefootnote{\fnsymbol{footnote}}
\setcounter{footnote}{0}

\section{Introduction}

Despite providing a very successful description of fundamental interactions at the electroweak (EW) scale, the Standard Model (SM) of particle physics still presents some significant inadequacies. Two notable examples are the absence of neutrino mass terms in the Lagrangian and the lack of a dynamical origin for the Higgs potential, that exposes it to possibly large radiative corrections in the presence of heavy states coupled to the Higgs field. 
Both issues typically require the SM to be extended in the UV with the introduction of new fields and symmetries.

In this talk we explore a scenario in which both neutrino and Higgs masses originate from the same new physics sector: the type I seesaw model.
This minimal scenario, that we refer to as the ``Neutrino Option''~\cite{Brivio:2017dfq,Brivio:2018rzm}, brings an interesting change in perspective compared to the traditional approaches to the hierarchy problem, that could be also implemented within different UV assumptions.

\section{The hierarchy problem in the EFT language}
We work in the theoretical framework of Effective Field Theories (EFTs), with the implicit assumption that heavy states beyond the SM can be safely integrated out of the spectrum when considering observables defined at the EW scale or lower.

In a theory containing a generic state with mass $m\gg v$  coupled to the Higgs doublet $H$, radiative corrections to $(H^\dag H)$ are typically induced, proportional to $m^2$. In the EFT where the heavy state is absent, these corrections are matched by finite {\it threshold matching contributions}.
The hierarchy problem can thus be expressed in terms of threshold contributions $\Delta m_h^2\gg (\unit[125]{GeV})^{2}$ being allowed in the presence of new heavy particles.

Traditional solutions invoke new symmetries in order to either enforce cancellations among threshold contributions (as in supersymmetry) or suppress them down to $\Delta m_h^2 \lesssim (\unit[125]{GeV})^{2}$ (as in composite Higgs models). Both these approaches typically stabilize the Higgs potential at the TeV scale, which can be problematic for two main reasons. On the one hand, the existence of new particles at this mass scale is not indicated by current measurements. On the other hand, generating the correct potential at this scale always requires a certain amount of tuning. In the specific case of composite Higgs models, this is expressed by the requirement $(v/f)^2< 1$, where $f$ is the compositeness scale~\cite{Bellazzini:2014yua}. Alternative approaches are therefore worth exploring. 

\vskip 5mm

\begin{SCfigure}[1][t]\centering
\includegraphics[width=9.2cm]{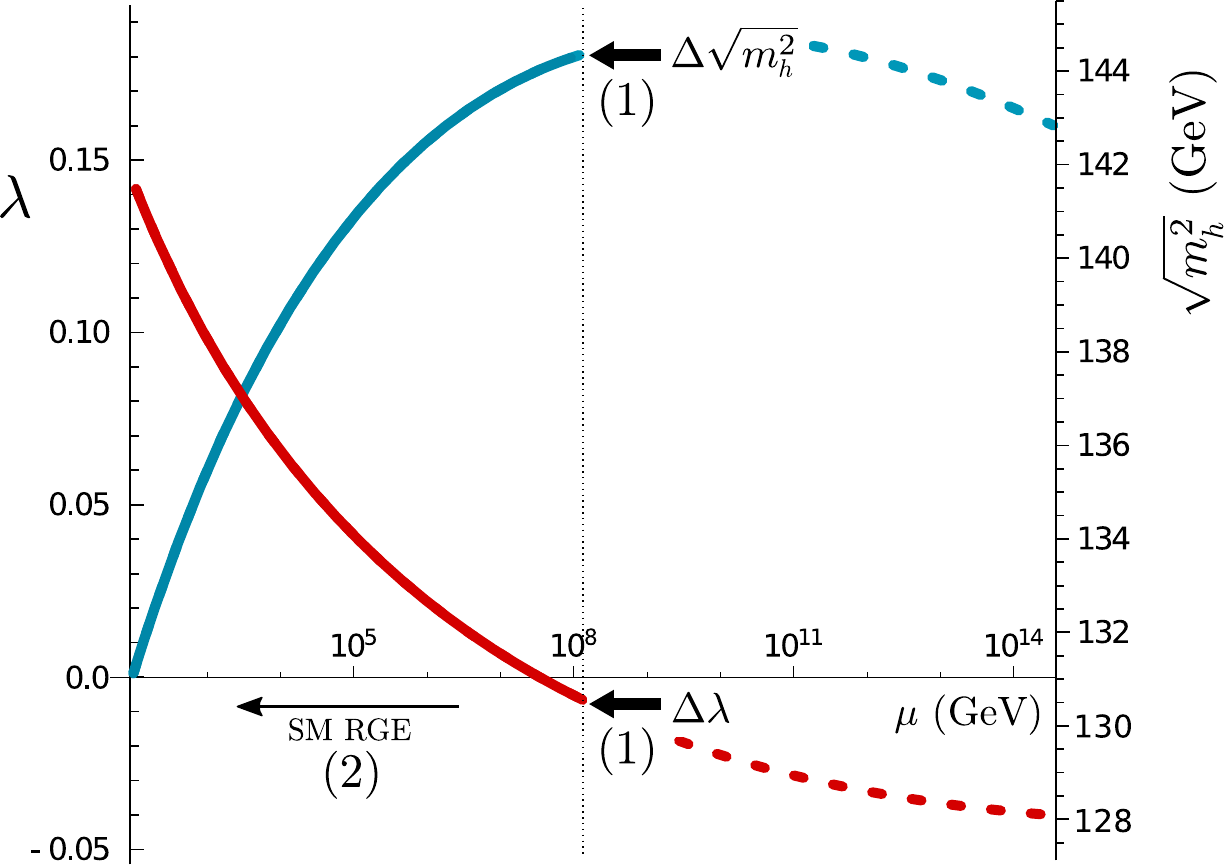}
\caption{ Schematic illustration of the main idea underlying the Neutrino\\ Option: (1) the Higgs potential is\\ generated by new physics states at a scale $M\gg \unit{TeV}$ ($M=\unit[10^8]{GeV}$ in the figure). (2) At $E<M$ the Higgs parameters run according to the SM RGEs, with boundary conditions fixed by the threshold matching contributions $\Delta \sqrt{m^2_h}$, $\Delta \lambda$. The figure shows in two overlapping plots the 1-loop SM running of the Higgs mass (blue line, right axis) and of the quartic coupling $\lambda$ (red line, left axis) for a top quark mass $\mt=\unit[173.2]{GeV}$~\cite{Buttazzo:2013uya}.   }\label{fig:running_schematic}
\end{SCfigure}

The main idea underlying the Neutrino Option is that the generation of the potential can be moved to a scale $M\gg $~TeV~\cite{Brivio:2017dfq}. 
At $E<M$ the new sector can be integrated out, and the Higgs mass and quartic coupling run down to the EW scale according to the the SM renormalization group equations (RGE)~\cite{Buttazzo:2013uya}, with boundary conditions fixed at $\mu=M$ by the threshold matching contributions.
This principle is illustrated in Figure~\ref{fig:running_schematic}, with the Higgs potential parameterized as
\begin{equation}\label{eq.V}
V(H) = -\frac{m_h^2}{2}H^\dag H + \lambda (H^\dag H)^2\,.
\end{equation}

Here it was implicitly assumed that the classical potential is approximately vanishing at the scale $M$, so that the threshold contributions dominate the RGE boundary conditions and ultimately control the vacuum structure. This assumption can be partially relaxed, for instance allowing a bare term $\lambda_0 (H^\dag H)^2$ with $\lambda_0>0$, which is compatible for instance with an approximate conformal symmetry in the UV.

The approach proposed here has the advantage that
constraints from direct searches of new particles are easily evaded. More importantly, it does not require the introduction of a complex new physics sector, as $\Delta m_h^2$ is easily induced in simple UV completions of the SM. A very interesting case is that of type I seesaw, whose primary motivation is to generate light neutrino masses. 
Below we show how this minimal extension of the SM can successfully generate the SM Higgs potential.

\section{The Neutrino Option}

We consider a standard type-I seesaw scenario~\cite{Seesaw}
, extending the SM  with $N$ singlet right-handed fields $N_{R,p}$, $\,p=\{1,\dots N\}$. In the basis in which the Majorana mass matrix $M_p$ is real and diagonal, the relevant Lagrangian is (we use the notation of~\cite{Broncano:2002rw,Elgaard-Clausen:2017xkq})
\begin{equation}
\L_{N_p}=  \frac12\overline{N_p} (i\slashed{\partial} - M_{p})N_p -
\left[\overline{\ell_{L}^\beta} \tilde{H} \omega^{p,\dagger}_\beta  N_p
+ \overline{N_p}\, \omega^p_\beta \tilde{H}^\dagger \ell_{L}^\beta
\right]\,.
 \label{LN}
\end{equation}
Here $\ell$ is the SM lepton doublet, with flavor index $\beta$. 
The fields $N_p$ satisfy the Majorana condition $N_p^c=N_p$, where the $c$ superscript denotes
charge conjugation for Dirac spinors, defined as $\psi^c = -i\g_2\g_0 \bar\psi^T$.
They are related to the chirality eigenstates by {$N_p = e^{i\theta_p/2} N_{R,p}+ e^{-i\theta_p/2}(N_{R,p})^c$}, being $\theta_p$ the Majorana phases~\cite{Broncano:2002rw}.
The couplings $\omega_\beta^p$ are complex $3\times N$ matrices in flavor space. Finally, here and in the following, repeated indices are summed over, unless otherwise stated.

Integrating out the  $N_p$ states and matching at treel-level onto the SM EFT gives the Majorana mass matrix for the 3 light neutrinos  
\begin{equation}\label{eq.m_nu}
(m_\nu)_{\a\beta} =-\frac{v^2}{2}(C_5)_{\a\beta} =
-\frac{v^2}{2}\frac{ \omega_\a^p \omega_\beta^p }{M_p}\,,
\end{equation}
with $C_5$ the Wilson coefficient of the dimemsion~5 Weinberg operator.
The matrix $m_\nu$ is diagonalized by a unitary rotation matrix $U_\nu$  
that, in the mass eigenbasis of the charged leptons, coincides with the Pontecorvo-Maki-Nakagawa-Sakata (PMNS) matrix~\cite{PMNS}, 
  customarily parameterized as
\begin{equation}\label{eq.PMNS_param}
U_{PMNS}  =
\begin{pmatrix}
c_{12} c_{13}  &    s_{12}c_{13}    &   s_{13} e^{-i\delta}\\ 
-s_{12}c_{23}-s_{23}c_{12}s_{13}e^{i\delta}   &    c_{12}c_{23} - s_{23}s_{12} s_{13} e^{i\delta}  &   s_{23}c_{13} \\
s_{12} s_{23} - c_{23}c_{12} s_{13} e^{i\delta}   &  -s_{23}c_{12}-c_{23}s_{12}s_{13} e^{i\delta}  &  c_{23} c_{13}
\end{pmatrix}
\begin{pmatrix}
e^{-i \phi /2}& & \\
& e^{-i \phi'/2} & \\
& & 1
\end{pmatrix}\,,
\end{equation}
where $c_{ij} = \cos\theta_{ij}, \, s_{ij}=\sin\theta_{ij}$. $\delta$ is the CP violating phase and $\phi,\,\phi'$ are the Majorana phases. The mixing angle and CP phase, together with two squared mass splittings $\Delta m^2_{ij} = m_i^2-m_j^2$,  can be constrained by neutrino oscillation measurements. In this analysis we consider the best fit values and $3\s$-allowed ranges of Ref.~\cite{Esteban:2016qun}.

Neglecting fermion masses of the first two generations and working in the charged leptons and quarks mass eigenbases, the RGE of $(C_5)_{\a\beta}$ can be compactly written~\cite{Babu:1993qv,Antusch:2001ck}
\begin{equation}
16\pi^2 \mu \frac{d C_5}{d\mu} = 
-\frac32\left[C_5 \cdot \diag(0,0,y_\tau^2) + \diag(0,0,y_\tau^2)\cdot C_5\right]
-\left[3g_2^2-4\lambda - 6 y_t^2-6 y_b^2-2y_\tau^2\right]C_5\,,
\end{equation}
with $g_2$ the $SU(2)$ coupling constant. This running can be equivalently described in terms of individual RGEs for each relevant parameter, as derived in Ref.~\cite{Casas:1999tg} (see also~\cite{Brivio:2018rzm}).

\begin{SCfigure}[.8][t]\centering
 \hspace*{5mm}\includegraphics[width=.73\textwidth]{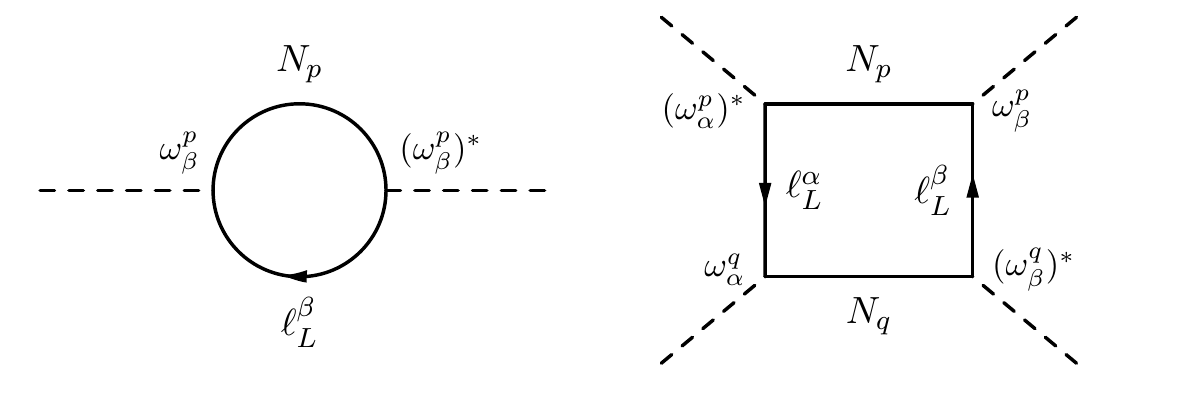}
 \caption{ One-loop diagrams leading to threshold matching contributions to the Higgs potential in the type-I seesaw model.}\label{fig:diagrams}
\end{SCfigure}

\vskip 5mm
Threshold corrections to the Higgs potential are induced at order one-loop in the seesaw model due to the diagrams in Fig.~\ref{fig:diagrams}. The finite terms are\footnote{The loops are calculated using dimensional regularization and $\overline{\rm MS}$ renormalization. The threshold corrections are then identified as the zero-th order term in the expansion in $v/M$. Finally, $\bar \mu = \mu_{\overline{\rm MS}} e^{3/4}$.}~\cite{Brivio:2017dfq,Brivio:2018rzm}
\begin{align}
\label{eq.Deltam2}
\Delta \mh^2 &=
\frac{M_p^2 |\omega_p|^2}{4 \, \pi^2}\,\Bigg[ \dfrac{1}{2} - \log \frac{\bar\mu^2}{M_p^2}\Bigg]\,,\\[-2mm]
\label{eq.DeltaLam}
\Delta \lambda &=
\frac{(\omega\omega^\dagger)_{pq}(\omega\omega^\dagger)_{qp}}{16 \,\pi^2} \, \Bigg[-\frac{1}{2} + \frac{M_p^2
\log \frac{\bar\mu^2}{M_p^2}-M_q^2 \log \frac{\bar\mu^2}{M_q^2}}{(M_p^2-M_q^2)} \,\Bigg]
-\frac{\left((\omega\omega^\dagger)_{pq}\right)^2}{16 \,\pi^2}
\frac{M_p M_q}{M_p^2-M_q^2}\log\frac{M_p^2}{M_q^2}
\,,
\end{align}
consistent with the results in~\cite{Thres}. 
The RG running of these quantities is coupled to the running of the other SM parameters; here we solve the system of equations of Ref.~\cite{Buttazzo:2013uya}.
\section{Preliminary study}
A preliminary study of the viability of the Neutrino Option was presented in Ref.~\cite{Brivio:2017dfq}. The main goal of this analysis is to verify the existence of regions of the parameter space of the seesaw model where both the generated Higgs potential and neutrino masses are compatible with experimental observations.
The flavor structure of the model can be neglected at this stage:  we introduce $N=3$ nearly degenerate Majorana states $M_p \simeq M$ and we simplify the notation for neutrino Yukawa coupling assuming $|\omega^p|\simeq \omega$.  Finally, we take the classical Higgs potential to be nearly vanishing at $E\gtrsim M$: $m_{h,0},\lambda_0\sim 0$. 

\begin{figure}[t]\centering
 \includegraphics[width=.8\textwidth]{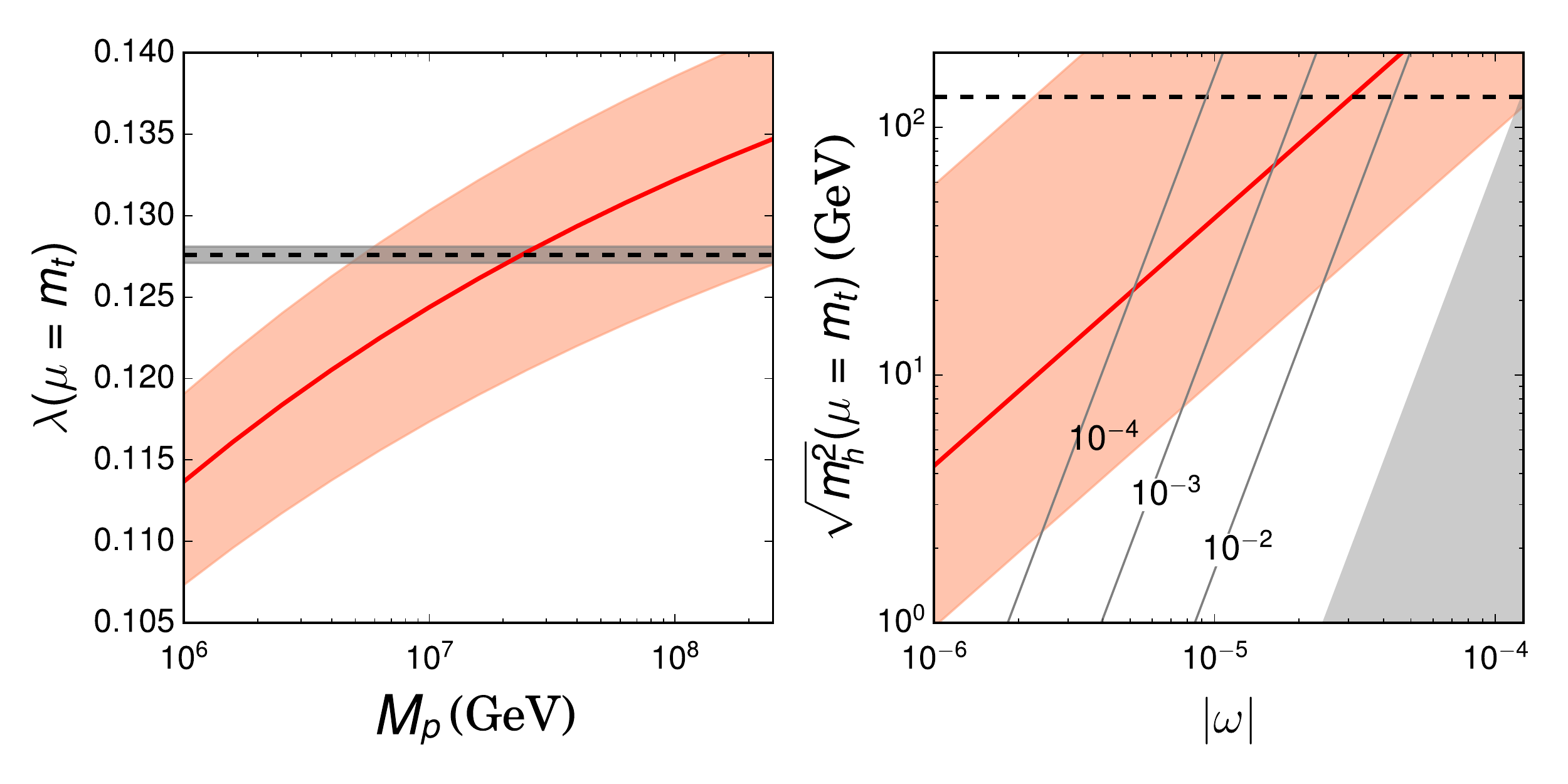}
 \caption{Values of the parameters $\lambda$ (left) and $\sqrt{\mh^2}$ (right) extrapolated at the scale $\mu=\mt$ as a function of the two seesaw parameters $M$ and $\omega$ respectively, in the preliminary study of Ref.~\cite{Brivio:2017dfq}. The dashed lines and surrounding bands indicate the values consistent with the measured Higgs mass within $\pm 1\sigma$~\cite{Aad:2015zhl}. Left panel: the red line assumes $\mt=173.2$~GeV and the orange band corresponds to varying $\mt$ between 171 and 175~GeV. Right panel:
the solid red line assumes $M = 10^{7.4}$~GeV. The grey region is disfavoured by the $\Lambda$ CDM cosmology limit $\sum m_\nu \leq 0.23$~eV. The three solid lines indicate, for reference, the sum of neutrino masses predicted, in ${\rm eV}$.}\label{fig.preliminary_plots}
\end{figure}

With this setup, there are only 2 relevant quantities in the seesaw parameter space, $M$ and $\omega$, and 3 quantities need to be matched for the Neutrino Option to be valid, namely the measured values of $\mh,\,\lambda,\, m_\nu$.
The threshold matching contributions of Eqs.~\leqn{eq.Deltam2},\leqn{eq.DeltaLam}, evaluated at $\bar\mu=M$, reduce to
\begin{equation}
\Delta \mh^2 = 3\,\frac{M^2 \omega^2}{8\pi^2},\qquad 
\Delta\lambda = -9\,\frac{5 \omega^4}{32\pi} \,.
\end{equation}
We extract the values of $M$ and $\omega$ for which the 1-loop RGE system with boundary conditions\footnote{
The boundary conditions for the relevant SM parameters $\{g_1,g_2,g_s,y_\tau,y_b,y_t\}$ are fixed at $\mu=\mt$. Because we use 1-loop RGEs, they are matched at tree level to their SM values. }
$\mh^2(M) = \Delta\mh^2$, $\lambda(M)=\Delta\lambda$ leads to the correct values of $\mh(\mt)$, $\lambda(\mt)$. Figure~\ref{fig.preliminary_plots} shows the results obtained.
The matching of $\lambda$ is nearly insensitive to variations of $\omega$ as long as $\omega < 1$, so it can be directly solved for $M$. For a top quark mass $\mt=\unit[173.2]{GeV}$~\cite{Patrignani:2016xqp} it selects $M\simeq\unit[10^{7.4}]{TeV}$ (left plot). With this $M$ fixed, the correct Higgs mass is generated for $\omega \simeq 10^{-4.5}$ (right plot). The size of light neutrino masses is now fixed, and can serve as a test: one finds $\sum m_\nu \simeq 3\omega^2 v^2 / 2 M\simeq \unit[3\cdot 10^{-3}]{eV}$, which nicely lies in the ballpark allowed by neutrino oscillations and cosmology bounds.

Although approximated and affected by large uncertainties (see Fig.~\ref{fig.preliminary_plots}), this result is particularly interesting, as it proves that two very different mass scales, $\mh$ and $m_\nu$, can be reproduced at the same time within the simple seesaw framework.

\section{Improved study}
At least two aspects of the preliminary study in Ref.~\cite{Brivio:2017dfq} can be improved: (i) flavor indices must be taken into account in order to verify whether the peculiar neutrino flavor structure can be reproduced, (ii) higher-loop order RGEs should be employed to obtain more accurate estimates: the results of Ref.~\cite{Buttazzo:2013uya} allow to consider 3-loop RGEs for $\lambda$ and $\mh$.  Finally  a bare term $\lambda_0$ in the high energy classical potential can be allowed.

\begin{figure}[t]\centering
\includegraphics[width=7.5cm]{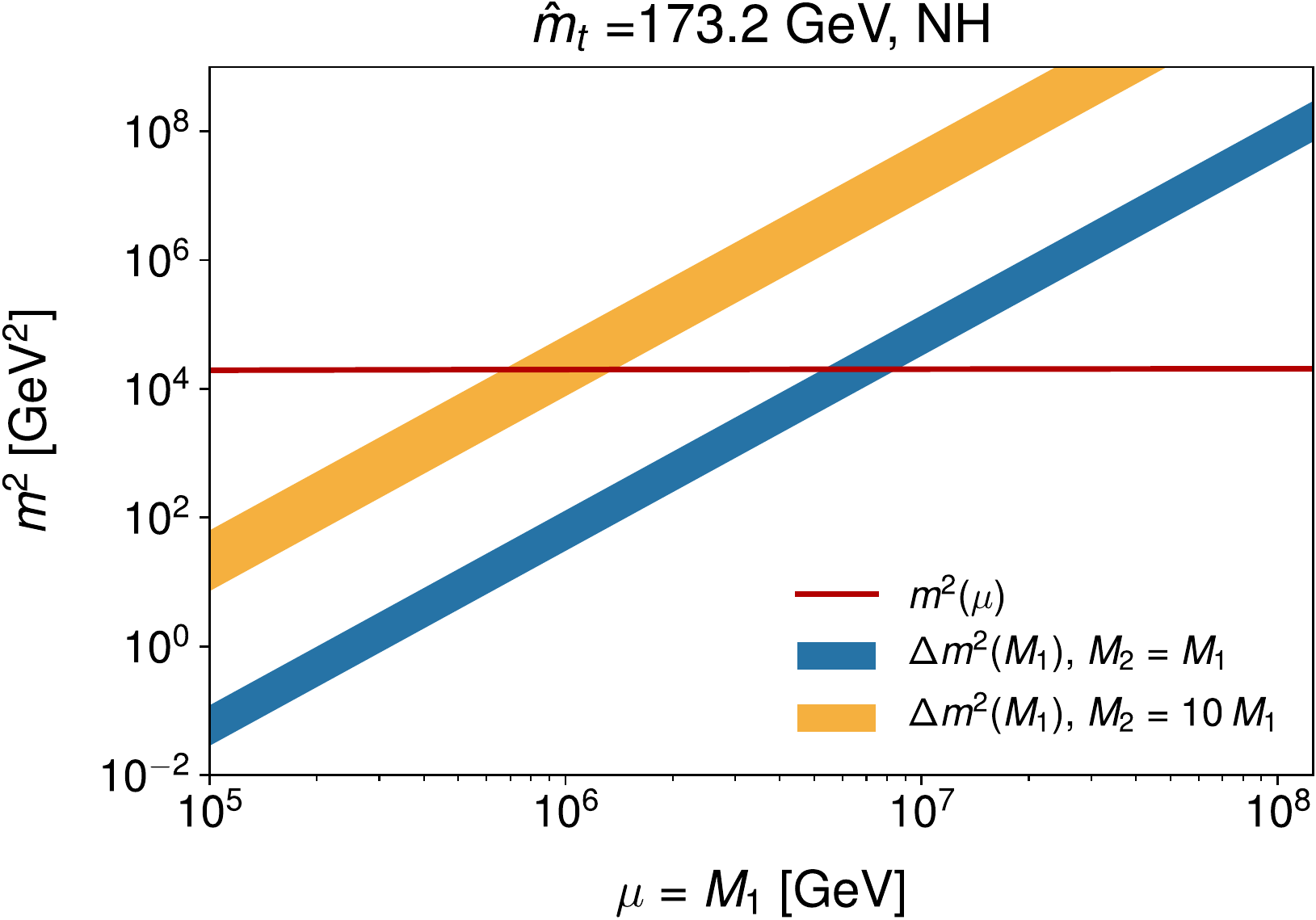}
\hfill
\includegraphics[width=7.5cm]{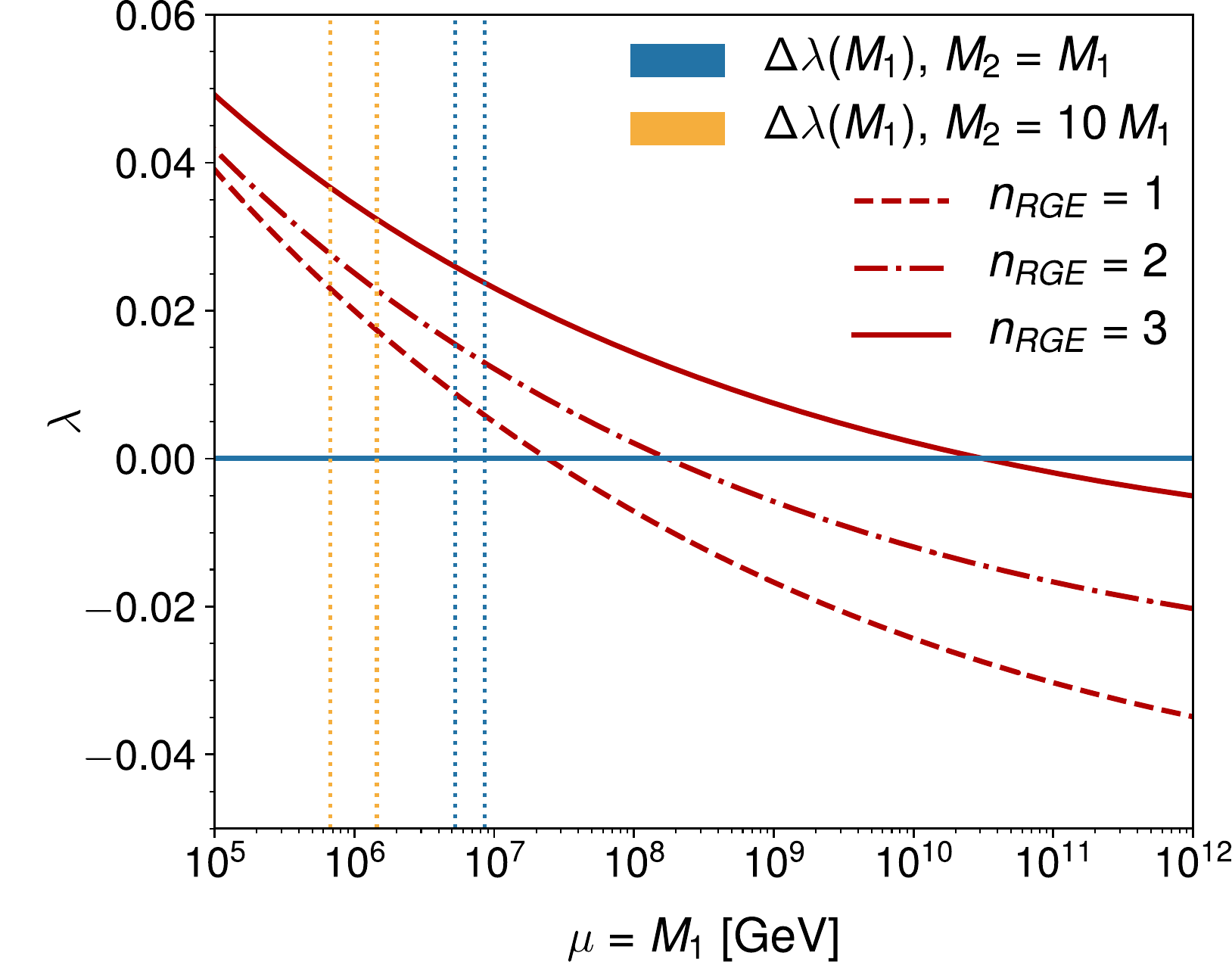}
\caption{Matching procedure in the refined analysis of Ref.~\cite{Brivio:2018rzm} for $\mh^2$ (left) and  $\lambda$ (right): the red curves show the running of the parameters determined by the SM RGEs at 1, 2 and 3 loops. 
The blue (yellow) band marks the values of the threshold matching contributions $\Delta\mh^2$ and $\Delta\lambda$ compatible at $3\s$ with neutrino physics constraints, for the benchmark $M_2=M_1$ ($M_2=10M_1$) and as a function of $M_1$. 
The red curves overlap in the left plot, and the bands overlap in the right plot. The dotted vertical lines mark the $M_1$ range where the matching for $\mh^2$ is fulfilled, for each benchmark. This plot assumes normal neutrino mass ordering and $\mt=\unit[173.2]{GeV}$.}\label{fig.refined_plots}
\end{figure}

\begin{figure}[t]\centering
\hspace*{-1cm}
\includegraphics[width=18cm]{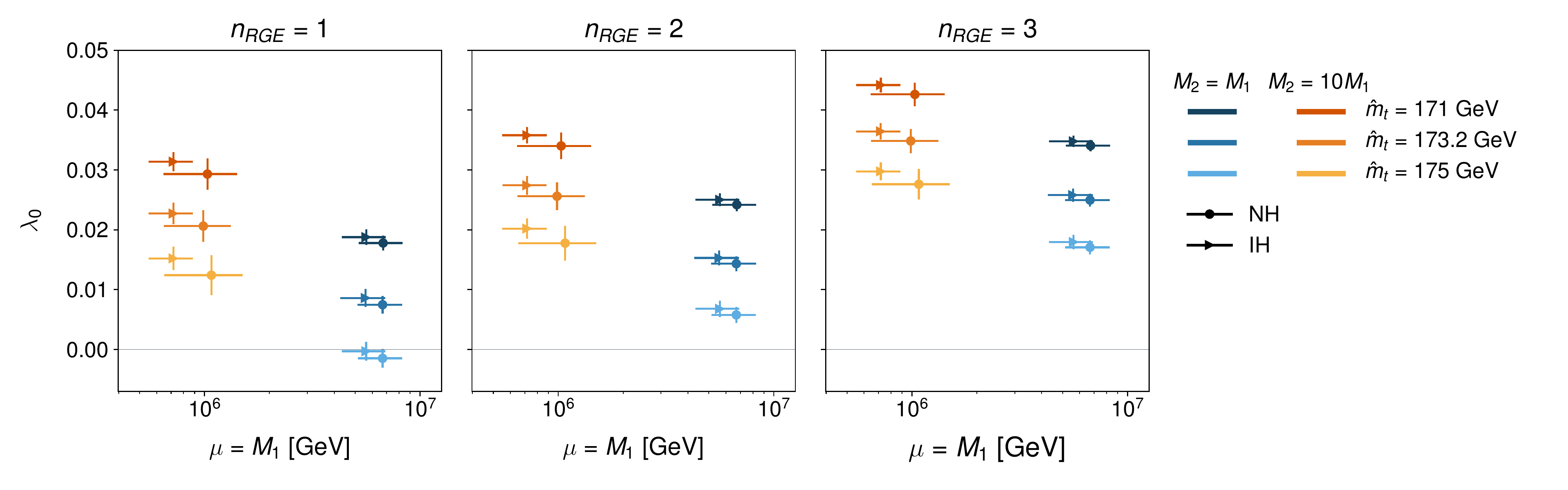}
\caption{Values of $(M_1,\lambda_0)$ for which the Neutrino Option can be realized compatibly with both Higgs and neutrino oscillation data, for different choices of: the SM RGE order (panels from left to right), the top mass $\mt$ (lighter to darker colors), the neutrino mass ordering (dot shapes) and $x_M$ (blue vs orange). The error bars reflect $3\s$ variations of the neutrino parameters being allowed in the scan.}\label{fig.summary}
\end{figure}

All these points have been addressed in Ref.~\cite{Brivio:2018rzm}: for simplicity here we introduce $N=2$ heavy Majorana states with masses $M_1\leq M_2$, so the lightest neutrino is exactly massless. 
In order to implement the flavor constraints and have stable numerics, we change the analysis strategy compared to Ref.~\cite{Brivio:2017dfq} and we  fix all the RGE boundary conditions at the lower scale: for the Higgs sector\footnote{The boundary conditions for the remaining SM parameters and for different choices of the RGE loop order are listed in Ref.~\cite{Brivio:2018rzm}.} we have 
$\lambda(\mt)_{2-{\rm loop}} =0.1258$,
$\mh(\mt)_{2-{\rm loop}} =\unit[131.431]{GeV}$\,.
For the neutrino sector we select the values of the parameters\footnote{Here $m_l=m_3$ for normal neutrino mass ordering and $m_l=m_1$ for inverted ordering.}
$\{m_2, m_l, \theta_{12},\theta_{13},\theta_{23},\delta,\phi,\phi'\}$ at $\mu=m_Z$ with a random scan of the $3\s$-allowed space from Ref.~\cite{Esteban:2016qun}.
For each point in the scan we run all the parameters up to an arbitrary scale $M_1>\mt$, where the full matrix $\omega_\beta^p(\mu=M_1)$ can be inferred via the Casas-Ibarra parameterization~\cite{Casas:2001sr} as a function of the neutrino parameters and of an additional complex quantity $r$, taken here to be a random number with $|r|\leq 1$. If $x_M = M_2/M_1$ is also fixed, the threshold matching contributions 
\begin{align}
\Delta\mh^2 &= \frac{M_1^2}{8\pi^2}\left(|\omega_1|^2 + x_M |\omega_2|^2\right]\,,\\
\Delta\lambda &=
-\frac{1}{32\pi^2}\left[5|\omega_1|^4 + 5|\omega_2|^4 + 2\im(\omega_1\cdot\omega_2^*)^2\left(1-\frac{2 \log x_M^2}{1+x_M}\right)
+ 2\re(\omega_1\cdot\omega_2^*)^2\left(1-\frac{2\log x_M^2}{1-x_M}\right)\right]\,,\nonumber
\end{align}
are a function of $M_1$ only and can be compared to  $\mh(\mu=M_1)$, $\lambda(\mu=M_1)$ extrapolated from the SM running.
This is illustrated in Figure~\ref{fig.refined_plots}, taking two benchmark cases $x_M=1$ and $x_M=10$. 
The Neutrino Option is realized when simultaneously
\begin{equation}
\mh^2 (M_1) = \Delta \mh^2\,,
\quad\text{ and }\quad
\lambda(M_1) = \Delta \lambda + \lambda_0\,.
\end{equation}
The values of $M_1$ and $\lambda_0$ that satisfy this condition for several sets of assumptions are reported in Figure~\ref{fig.summary}.

Because neither the neutrino parameters nor the Higgs mass run significantly, the $M_1$ range identified  is very stable under variations of $\mt$ and choice of the RGE order. 
In this sense it is a clean {\it prediction} of this scenario, consistent with a simple general estimate: requiring $M^2 \omega^2 /8\pi^2 \sim\mh^2\sim (\unit[100]{GeV})^2$ and $v^2\omega^2/2M\sim m_\nu \gtrsim \unit[0.01]{eV}$ one easily finds $\omega\sim \unit[1]{TeV}/M$ and $M\lesssim\unit[10^7]{GeV} = \unit[10]{PeV}$. 
Within this range, the $\lambda$ running curve cannot be matched by the threshold correction alone, as $\lambda(\mu=M_1)\gg \Delta\lambda\simeq 0$. A bare $\lambda_0\simeq 0.01 -0.05$ is therefore required for the Neutrino Option to be realized. Note that this value is sufficiently small for $\lambda_0$ to be radiatively generated  without further tuning in a putative UV completion.

\section{Conclusions and Outlook}
The two analyses presented in this talk show that the Higgs mass can be generated alongside neutrino masses and mixings within a very minimal extension of the SM: type-I seesaw. If two nearly-degenerate Majorana states with mass $M$ are introduced, this requires 
$5 \cdot 10^5 \lesssim M\lesssim \unit[10^7]{GeV}$, Yukawa couplings  $\omega\sim \unit[1]{TeV}/M$ and a quartic coupling  $\lambda_0\sim0.01-0.05$ in the high energy classical potential. 

In this interesting scenario, the EW symmetry breaking
is ultimately tied to the breaking of an approximate conformal symmetry in the UV and to the violation of the lepton number.
These two aspects are therefore worth exploring in connection to the Neutrino Option. For instance, Refs.~\cite{Brdar:2018vjq,Brdar:2018num} suggested a minimal, conformal UV completion that extends the SM with two extra scalar fields in order to generate dynamically the heavy Majorana scale via the Gildener-Weinberg mechanism. The authors showed that this construction can successfully realize the Neutrino Option, and at the same time avoid Landau poles up to the Planck scale, in significant regions of the parameters space. Moreover, for a subset of the latter, a strong first order phase transition can be realized, leading to observable gravitational waves signals~\cite{Brdar:2018num}.

Finally, it is interesting to investigate whether the Neutrino Option is compatible with the leptogenesis process. Thermal leptogenesis typically prefers values of the neutrino Yukawa coupling that lie outside the range selected by the Neutrino Option~\cite{Davoudiasl:2014pya} and is therefore disfavored in this case. However, resonant leptogenesis~\cite{Pilaftsis:2003gt} can take place in the presence of nearly degenerate Majorana states and it can be shown to be successful within the Neutrino Option parameter space~\cite{inPrep}.

\providecommand{\href}[2]{#2}\begingroup\raggedright\endgroup

\end{document}

%% file: econfmacros.tex
\def\beq{\begin{equation}}
\def\eeq#1{\label{#1}\end{equation}}
\def\eeqn{\end{equation}}

\def\beqa{\begin{eqnarray}}
\def\eeqa#1{\label{#1}\end{eqnarray}}
\def\eeqan{\end{eqnarray}}

\def\leqn#1{(\ref{#1})}

\let\bar=\overbar

\def\etal{{\it et al.}}

\def\eg{{\it e.g.}}

\def\L{{\cal L}}

\def\Dslash{\not{\hbox{\kern-4pt $D$}}}
\def\dslash{\not{\hbox{\kern-2pt $\del$}}}

\def\mt{m_t}
\def\g{\gamma}
\def\a{\alpha}

\def\s{\sigma}

\def\mh{m_h}

\def\msb{{\bar{\ssstyle M \kern -1pt S}}}

\def\diag{{\rm diag}}
\def\re{{\rm Re}}
\def\im{{\rm Im}}

